\documentclass{article}

\usepackage{epsfig,verbatim}

\title{\bf CONVERGENCE TO THE CRITICAL ATTRACTOR AT INFINITE 
AND TANGENT BIFURCATION POINTS.}

\author{R. TONELLI  \\
INFM-SLACS laboratory, 
INFN,
Physics Department, \\University of Cagliari,
I-09042 Monserrato, Italy
}

\begin{document}
\baselineskip=1.0cm 
\maketitle
\newcommand{\be}{\begin{equation}}
\newcommand{\ee}{\end{equation}}

\begin{abstract}
\baselineskip=1.0cm 
The dynamics of the convergence to the critical attractor for the logistic map 
is investigated. 
At the border of chaos, when the Liapunov exponent is zero, 
the use of the non-extensive statistical mechanics formalism allows to define 
a weak sensitivity or insensitivity to initial conditions. 
Using this formalism 
we analyse how a set of initial conditions spread all over the phase space 
converges to the critical attractor in the case of infinite bifurcation 
and tangent bifurcation points. We show that the phenomena is governed 
in both cases by a power-law regime but the critical exponents 
depend on the type of bifurcation and may also depend on the numerical experiment 
set-up. Differences and similarities between the two cases are also discussed. 
\end{abstract} 
\newpage
\section{Introduction} 
Nonlinear dissipative maps are widely used in the description 
of a variety of systems exhibiting complex behaviour, given their 
suitability for numerical simulations. Sensitivity to initial conditions 
is easy to investigate in association to the dynamics onto 
the fractal attractor. Physical properties like entropies and 
relaxation phenomena are easily computed and generalised to more 
sophisticated systems. 
The logistic map is the most famous and used among 
the nonlinear maps. Nevertheless it still presents some aspect of the dynamics 
not well understood or exploited. This is the case for special points 
of the map, like the border or onset of 
chaos, constituted by the infinite bifurcation 
points or the tangent bifurcation points, often indicated as 
{\em critical points}, where the Liapunov exponent vanishes and sensitivity 
to initial conditions is not exponential but polynomial in time. 
A vast literature 
[C. Tsallis et.al., 1997; F. Baldovin et al., 2002; 
U. Tirnakli et al., 1999; M. Coraddu et al., 2004] has recently tried to investigate 
these critical points especially in association to entropic 
or relaxation properties. The main feature of the logistic map 
at these points is a {\em power-law sensitivity or insensitivity} 
to initial conditions [P. Grassberger and M. Scheunert, 1981; 
C. Tsallis, 1988; F.A.B.F. De Moura et al., 2000], 
with consequent peculiar behaviour for the 
time evolution and the convergence to the attractor of an ensemble of points. 
Recently a power-law sensitivity to initial 
conditions has been rewritten  
[C. Tsallis et al., 1997; M.L. Lyra and C. Tsallis 1998] in the form 
\be
\xi(t) = [1+(1-q)\lambda_{q}t]^{1/(1-q)},
\label{sens}
\ee
where $\lambda_{q}$ defines the time scale after which the power-law behaviour 
begins. When used to investigate the expansion (weak sensitivity, 
namely 1/(1-q)$\geq$0) or contraction (weak insensitivity, conversely 
1/(1-q) $\leq$ 0) 
of a compact ensemble of points or grouped initial conditions 
this relation is connected to the evolution of the entropy of 
such a system. To be more clear let consider a partition
of the phase space of the logistic map in {\em W$_{box}$} cells, 
where we use the parameterisation 
\be
x_{n+1} = 1 - ax_{n}^{2}
\label{log}
\ee
so that the phase space is $-1 \leq x \leq +1$. One of the standard ways 
to define the entropy of a system is the Boltzmann-Gibbs (BG) formulation 
\be
S_{BG} = - \sum^{W_{box}}_{i=1} p_{i} \log(p_{i})
\ee
where $p_{i}$ defines the probability of occupation of the i-th cell as 
the ratio of points inside the cell with respect to the total number 
of points in the phase space. 
When the parameter {\em a}  assumes values such that the map behaves 
chaotically, the Liapunov exponent is positive and 
there is {\em strong sensitivity} to initial conditions. If one groups 
a set of points close together they will expand exponentially 
in time (at least for small time scales) and correspondingly 
the BG entropy will increase. It has been demonstrated [A.N. Kolmogorov, 1958; 
Ya. G. Sinai, 1959] that 
the rate of increase of the BG entropy, known as Kolmogorov-Sinai (KS) entropy 
and defined as
\be
S_{KS} = \lim_{t \to \infty} \frac{S(t) - S(0)}{t}
\label{KS}
\ee
equals the Liapunov 
exponent under certain conditions, giving the well known 
Pesin equality $\lambda = S_{KS}$ [Ya. Pesin, 1977], so that the 
entropy increases linearly with time.
Conversely, if the map behaves periodically, the Liapunov exponent 
is negative and the BG entropy will decrease. 

When the parameter {\em a} is set in a critical point the Liapunov 
exponent vanishes and a set of different features appears. 
Both at the onset of chaos and at tangent bifurcation points 
new definitions of the sensitivity and of the entropies are suitable 
to describe the weak sensitive/insensitive behaviour. The term 
weak sensitivity (insensitivity) designs a separation that increases 
(decreases) not exponentially (the Liapunov exponent is zero) but 
polynomially in time.

Two different numerical experiments present a notable interest 
with regard to the way the points evolve in time approaching the attractor. 

On the one extreme, it is interesting to choose (randomly or not), 
all the initial conditions in a single cell, so that the initial 
entropy is zero 
[E. Borges et al., 2002; M. Coraddu et al., 2004]. 
The number of cells W(t) occupied at time t then grows as:
\be
W(t) = [1+(1-q)K_{q}t]^{1/(1-q)}
\label{Wt}
\ee
where $K_{q}$ is the {\em generalised} KS entropy, defined 
as the rate of increase of the non-extensive generalised Tsallis entropy
\be
S^{T}_{q} = (1-\sum p^{q}_{i})/(q-1), 
\label{TsE}
\ee
appropriate to describe 
the behaviour of the logistic map 
at critical points [C. Tsallis 1988]. 

On the other extreme one can choose 
to spread the initial conditions all over the whole phase 
space, occupying most of (or all, if the set of initial points 
is at least equal to the number of cells) the partition cells 
[F.A.B.F. De Moura et al., 2000].
The fact that the parameter {\em a} has been set in a critical 
point allows the system to exhibit peculiar behaviour. 
The main interest of this paper consists in the numerical 
analysis of the time evolution of
a set of initial conditions spread in the whole phase space and the way 
they converge to the critical attractor. 

We analyse the case of 
weak insensitivity and make a comparison with the weak sensitivity case. 
Due to the criticality of the system, the temporal evolution 
may depend on the particular choice of the initial conditions. 
In our case, with initial conditions spread over the phase space, 
one expects 
a shrink toward the attractor
in both cases, namely {\em a} corresponding to the infinite bifurcation 
point ($a_{\infty} = 1.401155189$) and to the tangent bifurcation point
($a_{tg} = 1.75$). 

\section{The Convergence to the Critical Attractor}
We analyse the temporal evolution of an ensemble of N initial conditions 
randomly chosen and uniformly distributed in the whole phase space. 
Following the work done in [F.A.B.F. de Moura et al.]
we partitioned the phase space in W$_{box}$ equal cells and examined 
the number of occupied cells W(t) at time t for different  
ratios {\em r} = N/W$_{box}$. Physically {\em r} is an index 
giving the sampling ratio of the phase space.  
We used W$_{box}$ ranging from 5000 to 128000 
cells for the partition and ratios {\em r} varying from 
{\em r} = 0.1 to {\em r} = 10000, for both cases $a = a_{\infty} =  1.401155189 $ 
and $a = a_{tg} = 1.75 $. 
When the convergence to the critical attractor is 
analysed using the non-extensive Tsallis entropy 
(\ref{TsE})
, there is 
a particular value of the entropic index {\em q} 
[G.F.J. Ananos et al., 2004 and references therein] 
such that the entropy 
$S^{T}_{q}$ evolves in time at constant rate $K_{q}$, with 
\be
 K_{q} = \lim_{N \to \infty} [S_{q}(t) - S_{q}(0)]/t
\ee
In the case of relaxation onto the attractor, the number of occupied cells 
decreases in time giving a negative $K_{q}$. In the simplest case 
of equiprobability, so that each cell contains the same number 
of points, the relation (\ref{TsE}) gives 
\be 
S^{T}_{q} = \frac{W(t)^{1-q} -1 }{1-q}
\label{equi}
\ee
and the number of occupied cells is expected to present a power-law decay
\be
W(t) = [W(0)^{1-q} +  (1-q)K_{q}t]^{1/(1-q)}
\label{owlow}
\ee
with asymptotic exponent $\alpha$ = 1/(1-q) $\leq$ 0. 
\\
In order to exploit most of the features of the 
convergence to the critical 
attractor we set up a series of different numerical experiments 
varying W$_{box}$ at constant {\em r} and vice-versa, for the two cases 
$a = a_{\infty} $
and $a = a_{tg} $.
The numerical experiments consist in selecting, using a uniform random 
distribution, at time zero, N initial points in the phase space of the logistic 
map  x $\in$ [-1.1], that has been partitioned in W$_{box}$ boxes. 
As each of the N points is iterated in time according to the logistic map 
W(t) boxes, among all the W$_{box}$, will be not empty at time t. The number 
W(t) generally will decrease iterating the map and is the quantity 
used in the following to characterize the different regimes of convergence 
to the attractor of the N initial points. 

As a first case we analyse the infinite bifurcation point. 
Here the expected power-law behaviour (Eq.~(9)) [P. Grassberger and M. Scheunert, 1981]
sets up for $a = a_{\infty} $ even at low {\em r} values ({\em r} = 0.1);
increasing {\em r} a superimposed log-log oscillation appears clearly 
and becomes more 
evident at higher {\em r} values (Fig.~1). 
In these cases there is a transient behaviour, preceeding the power-law 
regime , which is longer for 
larger {\em r}. 

It is important to note that, in a 
logarithmic-to-logarithmic scale, 
the transient time $\tau_{crossover}$ after which the convergence starts is translated 
by almost the same amount when r increases of a factor 10. 
This feature seems to 
persist up to the values 
of {\em r} = 10000, reached in our experiments. 
Higher values for the sampling ratio are in order to be 
used to verify the power-law regime in the limit {\em r} $\to \infty $.

The dependence of $\tau_{crossover}$ on the samping ratio {\em r} 
is clearly evidenced in Fig.~1. 
Previous works analysed values of sampling ratio up 
to {\em r} = 10, obtaining as a result a crossover time $\tau_{crossover}$ 
independent on {\em r} and related to the Eq. (\ref{powlow}) by $\tau_{crossover} 
\sim \frac{1}{W(0)^{q-1}}$. 
Analysing sampling ratios up to {\em r} = 10000, for many different W$_{box}$, this 
picture breaks down revealing a net and regular variation of $\tau_{crossover}$ 
with {\em r}.

In our analysis the power-law exponent seems to be not independent on {\em r}. 
The relation 
\begin{equation}
W(t) = t^{-\mu}P(\frac{ln(t)}{ln(\lambda)}) 
\end{equation}
holds in the hypothesis of log-log oscillations 
[D. Sornette et al., 1998; C. Tsallis et al., 1997], 
where P() is a period one function and $\lambda$ is a scaling factor between 
two oscillation periods. If the exponent $\mu$ is constant a plot of W(t)/t$^{-\mu}$ 
should show a periodic function 
in log(t) with constant average. 
Figure~2 shows that the log-periodic oscillations exist 
but the function P() can be globally increasing or decreasing in average, 
depending 
on the value of {\em r}. This trend can be explained if the exponent $\mu$ is not 
perfectly constant but slightly varying with {\em r} (at least for 
the considered values; it may be possible that increasing {\em r} a limit value for $\mu$ 
shows up. This value should probably be less than -0.71, around -0.8). 
In the figure we used the value $\mu$ = -0.71 
according to [F.A.B.F. De Moura et al., 2000].
However to the same sampling ratio {\em r} corresponds the same exponent 
irrespectively of W$_{box}$.

In the case of the tangent bifurcation point 
the regime of convergence to the critical attractor 
also presents a power-law behaviour, but with different exponents (Fig.~3). 
Fitting the curves we found $\mu$ = -1.55 for {\em r} = 1000 and $\mu$ = -1.45 
for {\em r} = 10000 (Fig.~4a).
This difference is not surprising, since 
the exponent is related to the structure of the 
chaotic attractor [F.A.B.F. De Moura et al., 2000].
Nevertheless a power-law 
behaviour was not obvious at the tangent bifurcation point because the critical attractor 
has only three fixed point and does not posses a multifractal structure. 
To our knowledge this is the first study of such a power-law behaviour 
during the convergence to the attractor 
of a statistical ensemble of initial conditions in a critical 
point with not infinite bifurcations or multifractal structure. 
We found (Fig.~3), as in the case of the infinite bifurcation point (Fig.~1), 
small irregular oscillations perturbing the power-law behaviour 
(to not be confused with the log-log oscillations) due to the 
finite size partition of the phase space. 
The larger is the number of cells W$_{box}$ (the finer is the partition) the 
smaller are the oscillations around the power-law behaviour. 
These oscillations are irregular and grid (and/or {\em r}) depending, 
and are almost completely reduced when W$_{box}$ is big enough. 

No log-log 
oscillations are observed in this case (Fig. 4b), reflecting the different 
structure of the attractor. 

Again, increasing {\em r}, in the log-to-log scale, 
the transient time $\tau_{crossover}$ after which the power-law 
regime of convergence begins is translated 
by approximatively 
the same amount when {\em r} is increased of a factor 10 
(Fig.~3). The ratio W(t)/W(0) before 
$\tau_{crossover}$ is determined in both cases 
by the portion of phase space occupied by the 
attractor at the edge of chaos.

We repeated the experiments with a different kind of initial set-up, 
choosing all the initial N points concentrated in a single cell, 
for $a$ = $a_{\infty}$ and $a$ = $a_{tg}$, using as a test case 
a grid of 64000 cells. 
At the infinite bifurcation point we found that, after a transient time, 
the system reaches 
a power-law regime with a different exponent with respect to 
the experiment with 
all the initial conditions spread in x $\in$ [-1,1] (Fig.~5a), so 
that the exponent is sensitive to the system set-up, namely to its 
past history. 
This behaviour has already been observed in others numerical experiments
at the infinite bifurcation point [F. Baldovin and A. Robledo, 2002b; 
U. Tirnakli et al., 1999].

A main difference 
is observed 
in the case 
of the tangent bifurcation. 
We repeated the same experiment for $a$ = $a_{tg}$ 
with all the N initial conditions arranged in a 
single cell and let them expand entropically until the process of 
convergence to the attractor 
begins. After a transient time we found exactly the same power-law 
index as in the case of initial conditions uniformly distributed 
over the phase space (Fig.~5b). 
In this case the system seems to be not sensitive to the system set-up or to its
past history.

\section{Conclusions}
In this paper we numerically studied how a statistical ensemble of initial conditions 
converges to the critical attractor using the logistic map. 
We compared the behaviour in two selected critical points: $a_{\infty}$ 
and $a_{tg}$. The main results can be summarised 
in the following points: 
\newline
- for $a$ = $a_{\infty}$ a power-law regime with superimposed log-log periodic 
oscillations sets up after different crossover times, depending on the sampling 
ratio {\em r} and 
there is no strict numerical evidence for 
the power-law exponent to be a constant independent 
on the sampling ratio, at least for {\em r} $\leq$ 100. 
\newline
- for $a$ = $a_{tg}$ a power-law convergence to the critical attractor, 
with different exponent, appears after a transient time even if the attractor 
has not fractal structure. The crossover time depends on the sampling ratio 
{\em r} in the same fashion as in the $a_{\infty}$ case and is partition independent;
\newline
- the choice of an ensemble of initial conditions concentrated on a single cell 
presents the same power-law exponent in the $a_{tg}$ case, whilst a different 
regime of convergence to the critical attractor appears in the 
case $a$ = $a_{\infty}$.

\newpage
\section*{References.}

Ananos G.F.J. and Constantino Tsallis [2004] 
``Ensemble Averages and Nonextensivity at the Edge of Chaos of One-Dimensional Maps", 
Phys. Rev. Lett.  {\bf 93}, 020601.

Baldovin F., A. Robledo [2002], 
``Universal renormalization-group dynamics at the onset of chaos in logistic maps and nonextensive statistical mechanics", 
Phys. Rev. {\bf E66}, 045104.

Baldovin F., A. Robledo [2002], 
``Sensitivity to initial conditions at bifurcations in one-dimensional nonlinear maps: Rigorous nonextensive solutions ",
Europhys. Lett. {\bf 60}, 518-524.

Borghes E., C. Tsallis, G.F.J. Ananos, P.M.C. de Oliveira [2002]
``Nonequilibrium Probabilistic Dynamics of the Logistic Map at the Edge of Chaos", 
Phys. Rev. Lett. {\bf 89}, 254103.

Coraddu M., F. Meloni, G. Mezzorani and R. Tonelli [2004], 
``Weak insensitivity to initial conditions at the edge of chaos in the logistic map"
, Physica {\bf A340}, 234-239.

De Moura F.A.B.F., U. Tirnakli and M.L. Lyra [2000], 
``Convergence to the critical attractor of dissipative maps: Log-periodic oscillations, fractality, and nonextensivity", 
Phys. Rev. {\bf E62}, 6361-6365.

Grassberger P. and  Scheunert M. [1981], 
``Some More Universal Scaling Laws for Critical Mappings", 
J. Stat. Phys. {\bf 26}, 697.

Kolmogorov A.N. [1958], 
``A new metric invariant of transient dynamical systems and automorphisms
 in Lebesgue spaces", 
Dokl. Akad. Nauk. SSSR {\bf 119}, 861; Ya.G. Sinai [1959], 
``On the Concept of Entropy of a Dynamical System",
{\em ibid.} 
{\bf 124}, 768.

Lyra M.L. and C. Tsallis [1998], 
``Nonextensivity and Multifractality in Low-Dimensional Dissipative Systems",
Phys. Rev. Lett. {\bf 80}, 53-56.

Pesin Ya. [1977], 
``Characteristic Lyapunov exponents and smooth ergodic theory", 
Russ. Math. Surveys {\bf 32}, 55-114.

Sornette D. et al. [1996], 
``Complex Fractal Dimensions Describe the Hierarchical Structure of Diffusion-Limited-Aggregate Clusters", 
Phys. Rev. Lett. 76, 251-254.

Tirnakli U., C. Tsallis, and M.L. Lyra [1999], 
``Circular-like maps: sensitivity to the
 initial conditions, multifractality and nonextensivity", 
Eur. Phys. J. {\bf B11}, 309-315. 

Tsallis C., A.R. Plastino, and W.-M. Zheng [1997], 
``Power-law sensitivity to initial conditions-new entropic
 representation", 
Chaos, Solitons Fractals {\bf 8}, 885-891.
     
Tsallis C. [1988], 
``Possible Generalization of Boltzmann-Gibbs Statistics",
J. Stat. Phys. {\bf 52}, 479-487. 
		  
Tsallis C. et al. [1997], 
``Specific heat anomalies associated with Cantor-set energy spectra"
Phys. Rev. {\bf E56}, R4922-R4925.

\begin{figure}[htb]
\begin{minipage}[b]{1.0\linewidth}
\centering
\centerline{\epsfig{figure=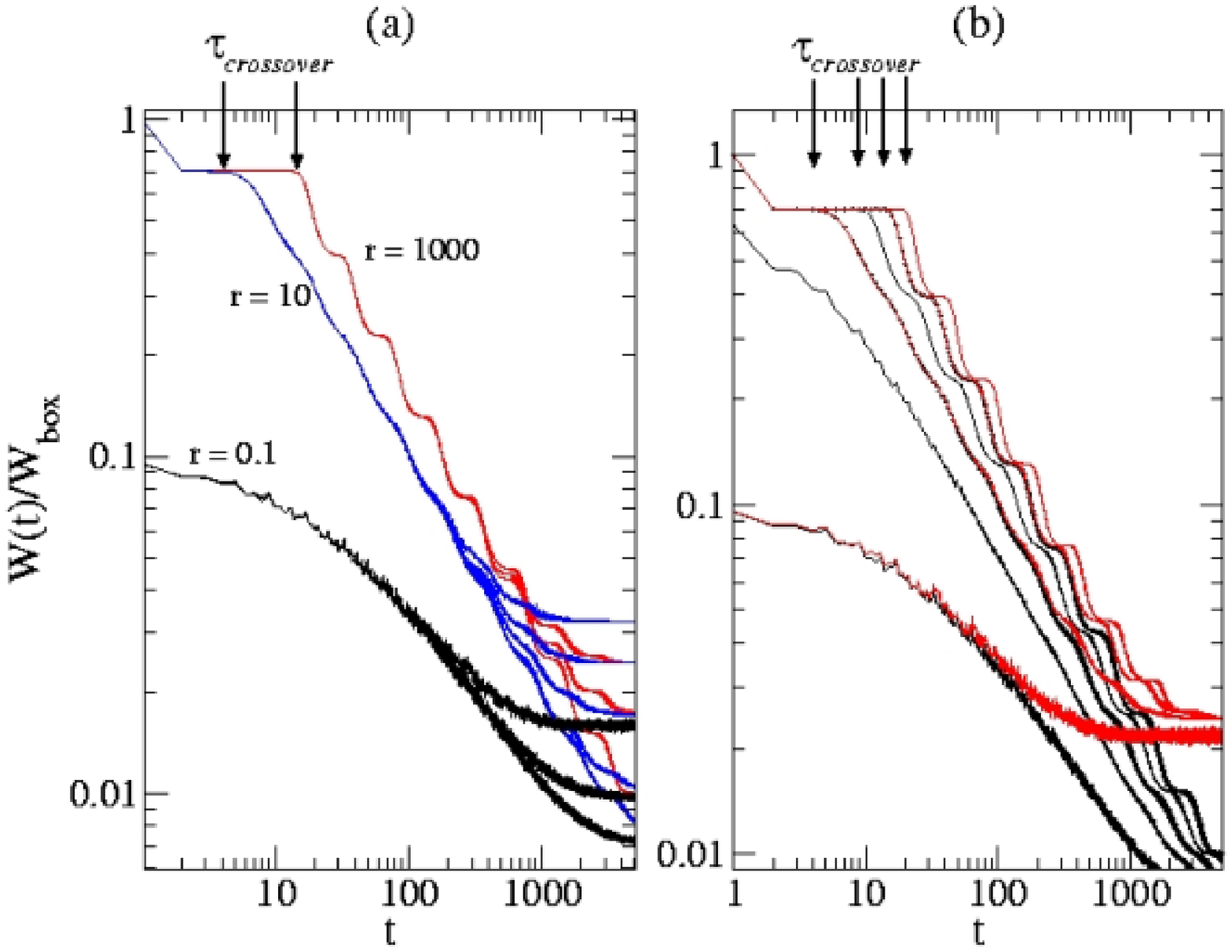,width=12.5cm}}
\end{minipage}
%
\newpage
\caption{
\baselineskip=1cm
We report for $a$ = $a_{\infty}$, in a log-log scale, the time 
evolution of the ratio between the occupied boxes and 
the available boxes. In frame (a) 
we show for different partitions $W_{box}$ 
of the interval [-1,1] the superimposed 
curves corresponding to the same {\em r}. We used the values 
$W_{box}$ = [20, 64, 128]$\times$ 10$^{3}$ for {\em r} = 0.1, 
$W_{box}$ = [5, 10, 20, 64, 128]$\times$ 10$^{3}$ 
for {\em r} = 10 and $W_{box}$ = [10, 20, 128]
$\times$ 10$^{3}$ for {\em r} = 1000. It is evident the 
occurrence of a power-law regime.
Superimposed log-log oscillations appear for higher sampling ratios. 
Crossover times are also indicated. To 
the same sampling ratio {\em r} corresponds the 
same power-law exponent for the various 
grids until saturation is reached at different times. In frame (b) we show 
in sequence the values {\em r} = 0.1, 1, 10, 100, 1000 and 10000 for the two grids 
$W_{box}$ = 10000 (red) and $W_{box}$ = 128000 (black). 
The indicated crossover times 
show a net dependence on the sampling ratio {\em r}.}
\label{fig:1}
\end{figure}
\newpage

\begin{figure}[htb]
\begin{minipage}[b]{1.0\linewidth}
\centering
\centerline{\epsfig{figure=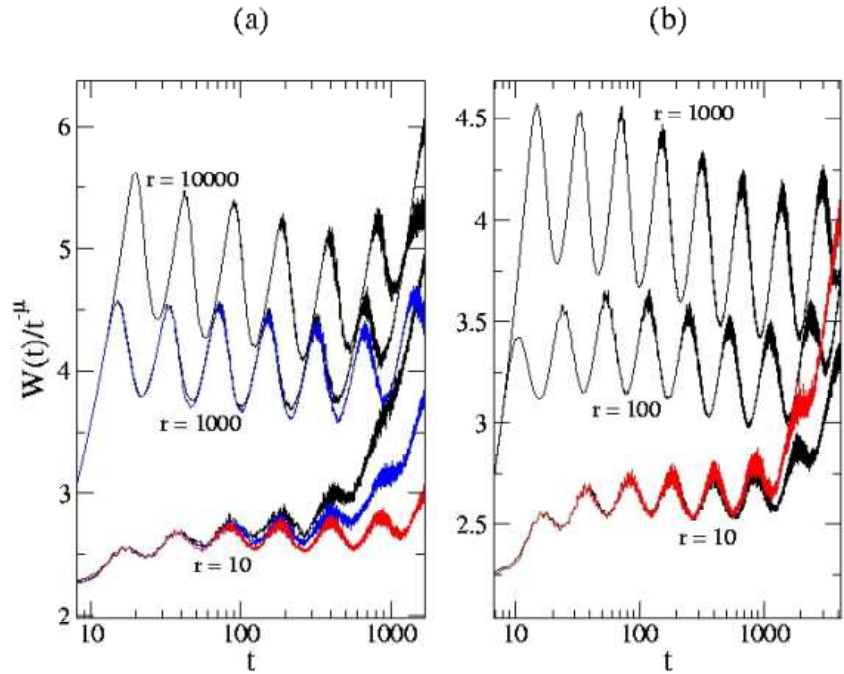,width=12.5cm}}
\end{minipage}
%
\newpage
\caption{
\baselineskip=1cm
We show the ratio W(t)/t$^{-0.71}$ versus log(t) for the infinite bifurcation 
point. In frame (a) we used $W_{box}$ = [10 (black), 20 (blue), 128 (red)]$\times$ 
10$^{3}$ for different sampling ratios {\em r}. Frame (b) shows different sampling 
ratios calculated using $W_{box}$ = 20000 (red) and $W_{box}$ = 128000 (black). 
Higher sampling ratios exhibit a decreasing mean value 
of W(t)/t$^{-0.71}$, while the lower ones exhibit an increasing mean value, 
suggesting a slight dependence of the power-law exponent on {\em r}. 
}
\label{fig:2}
\end{figure}

\newpage
\begin{figure}[htb]
\begin{minipage}[b]{1.0\linewidth}
\centering
\centerline{\epsfig{figure=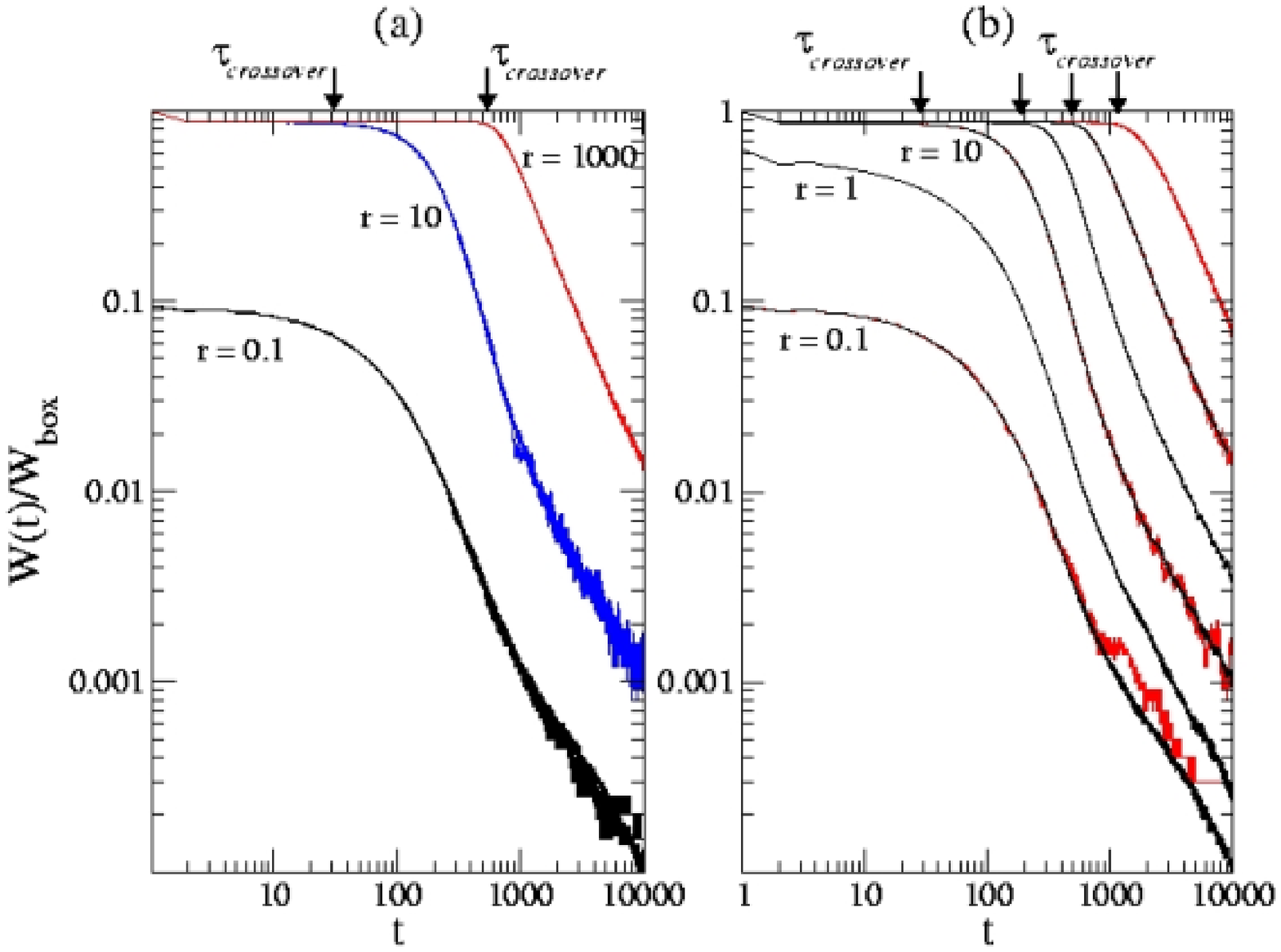,width=12.5cm}}
\end{minipage}
%
\newpage
\caption{
\baselineskip=1cm
We report for $a$ = $a_{tg}$, in a log-log scale, the time
evolution of the ratio of occupied boxes to the available boxes. In frame (a)
we show for different partitions of the interval [-1,1] the superimposed
curves corresponding to the same {\em r}. We used the values
$W_{box}$ = [20, 64, 128]$\times$ 10$^{3}$ for {\em r} = 0.1,
$W_{box}$ = [5, 10, 20, 64, 128]$\times$ 10$^{3}$
for {\em r} = 10 and $W_{box}$ = [10, 20, 128]
$\times$ 10$^{3}$ for {\em r} = 1000, as in the $a_{\infty}$ case. It is evident the
occurrence of a power-law regime well defined for higher sampling ratios 
with not superimposed log-log oscillations.
Crossover times are indicated. Again, to
the same sampling ratios {\em r} corresponds the
same power-law exponent for the various
grids. In frame (b) we show
in sequence the values {\em r} = 0.1, 1, 10, 100, 1000 and 10000 for the two grids
$W_{box}$ = 10000 (red) and $W_{box}$ = 128000 (black).
The indicated crossover times
show again a net dependence on the sampling ratio {\em r}. The power-law 
regime is well pronounced at higher {\em r} values and 
extends in time over different 
orders of magnitude. Saturation effects appear for lower grids.}
\label{fig:3}
\end{figure}

\newpage
\begin{figure}[htb]
\begin{minipage}[b]{1.0\linewidth}
\centering
\centerline{\epsfig{figure=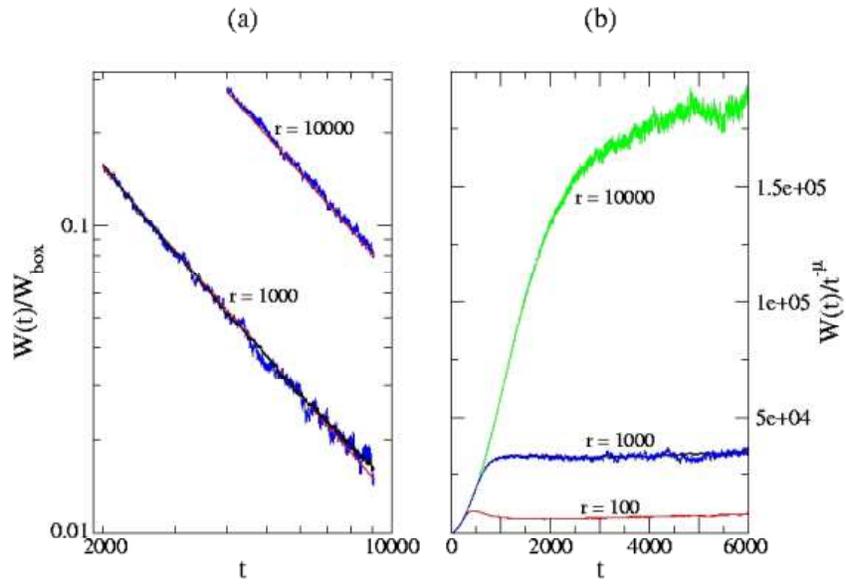,width=12.5cm}}
\end{minipage}
%
\newpage
\caption{
\baselineskip=1cm
Frame (a) evidences the power-law regimes for the tangent bifurcation point, 
with $W_{box}$ = 
[10 (blue), 128 (black)]$\times$ 10$^{3}$ and {\em r} = 1000, 
and with $W_{box}$ = 10000 and {\em r} = 10000 (blue). 
Red lines are fits to the curves, with exponents $\mu$ = -1.55 for {\em r} = 1000 and 
$\mu$ = -1.45 for {\em r} = 10000. Different exponents describe the power-law 
regime for different sampling ratios. The power-law 
persists long in time over different 
orders of magnitude.
Frame (b) shows the evolution of W(t)/t$^{-1.55}$ versus log(t) for various {\em r} 
and grids $W_{box}$ = 128000 (black) and $W_{box}$ = 10000 (other colours). 
The exponent $\mu$ = -1.55 is appropriate only for the case {\em r} = 1000 
where W(t)/t$^{-1.55}$ shows a constant mean value. 
}
\label{fig:4}
\end{figure}

\newpage
\begin{figure}[htb]
\begin{minipage}[b]{1.0\linewidth}
\centering
\centerline{\epsfig{figure=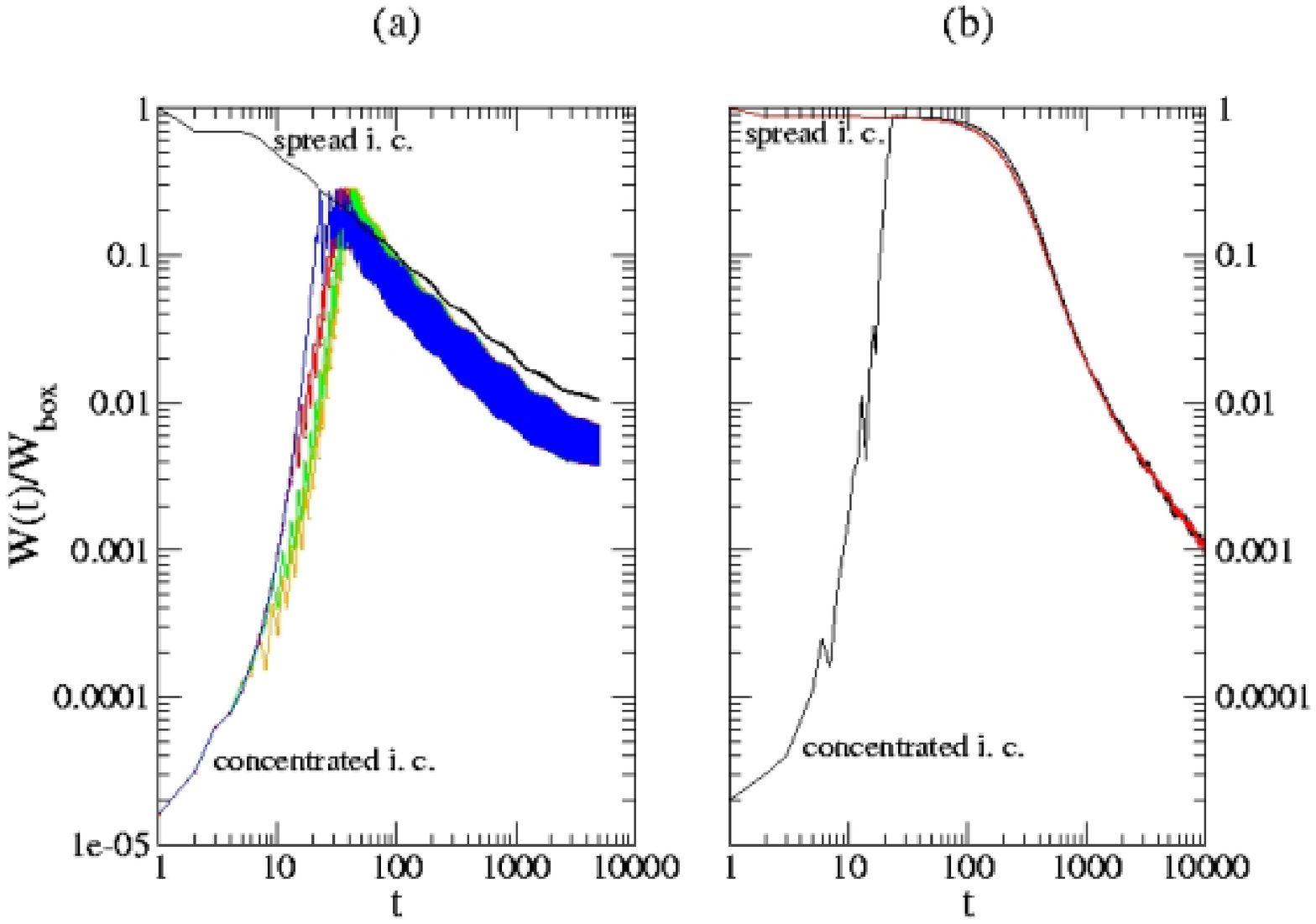,width=12.5cm}}
\end{minipage}
%
\newpage
\caption{
\baselineskip=1cm
We compare the different behaviours for $a = a_{\infty}$ (frame (a)) and 
$a = a_{tg}$ (frame (b)), when all the N initial conditions are randomly selected 
inside a single bin of the phase space partition. 
We used in both cases W$_{box}$ = 64000. 
Frame (a) illustrates the two 
different power-law coefficients obtained for $a = a_{\infty}$ when 
the initial conditions are uniformly spread all over the phase space or are 
constrained inside a single bin. 
In the frame we show four different selections of the initial bin in 
different colours. The four curves overlap in the regime of convergence to 
the attractor, oscillating at high frequency among two states every 
temporal step and giving rise 
to the broad coloured curves. 
In frame (b) the two different selections 
of initial conditions exhibit the same power-law exponent.
}
\label{fig:5}
\end{figure}

\newpage

{
\baselineskip=1cm
Figure 1\\
We report for $a$ = $a_{\infty}$, in a log-log scale, the time 
evolution of the ratio between the occupied boxes and 
the available boxes. In frame (a) 
we show for different partitions $W_{box}$ 
of the interval [-1,1] the superimposed 
curves corresponding to the same {\em r}. We used the values 
$W_{box}$ = [20, 64, 128]$\times$ 10$^{3}$ for {\em r} = 0.1, 
$W_{box}$ = [5, 10, 20, 64, 128]$\times$ 10$^{3}$ 
for {\em r} = 10 and $W_{box}$ = [10, 20, 128]
$\times$ 10$^{3}$ for {\em r} = 1000. It is evident the 
occurrence of a power-law regime.
Superimposed log-log oscillations appear for higher sampling ratios. 
Crossover times are also indicated. To 
the same sampling ratio {\em r} corresponds the 
same power-law exponent for the various 
grids until saturation is reached at different times. In frame (b) we show 
in sequence the values {\em r} = 0.1, 1, 10, 100, 1000 and 10000 for the two grids 
$W_{box}$ = 10000 (red) and $W_{box}$ = 128000 (black). 
The indicated crossover times 
show a net dependence on the sampling ratio {\em r}.}

\newpage
{
\baselineskip=1cm
Figure 2\\
We show the ratio W(t)/t$^{-0.71}$ versus log(t) for the infinite bifurcation 
point. In frame (a) we used $W_{box}$ = [10 (black), 20 (blue), 128 (red)]$\times$ 
10$^{3}$ for different sampling ratios {\em r}. Frame (b) shows different sampling 
ratios calculated using $W_{box}$ = 20000 (red) and $W_{box}$ = 128000 (black). 
Higher sampling ratios exhibit a decreasing mean value 
of W(t)/t$^{-0.71}$, while the lower ones exhibit an increasing mean value, 
suggesting a slight dependence of the power-law exponent on {\em r}. 
}

\newpage
{
\baselineskip=1cm
Figure 3\\
We report for $a$ = $a_{tg}$, in a log-log scale, the time
evolution of the ratio of occupied boxes to the available boxes. In frame (a)
we show for different partitions of the interval [-1,1] the superimposed
curves corresponding to the same {\em r}. We used the values
$W_{box}$ = [20, 64, 128]$\times$ 10$^{3}$ for {\em r} = 0.1,
$W_{box}$ = [5, 10, 20, 64, 128]$\times$ 10$^{3}$
for {\em r} = 10 and $W_{box}$ = [10, 20, 128]
$\times$ 10$^{3}$ for {\em r} = 1000, as in the $a_{\infty}$ case. It is evident the
occurrence of a power-law regime well defined for higher sampling ratios 
with not superimposed log-log oscillations.
Crossover times are indicated. Again, to
the same sampling ratios {\em r} corresponds the
same power-law exponent for the various
grids. In frame (b) we show
in sequence the values {\em r} = 0.1, 1, 10, 100, 1000 and 10000 for the two grids
$W_{box}$ = 10000 (red) and $W_{box}$ = 128000 (black).
The indicated crossover times
show again a net dependence on the sampling ratio {\em r}. The power-law 
regime is well pronounced at higher {\em r} values and 
extends in time over different 
orders of magnitude. Saturation effects appear for lower grids.}

\newpage
{
\baselineskip=1cm
Figure 4\\
Frame (a) evidences the power-law regimes for the tangent bifurcation point, 
with $W_{box}$ = 
[10 (blue), 128 (black)]$\times$ 10$^{3}$ and {\em r} = 1000, 
and with $W_{box}$ = 10000 and {\em r} = 10000 (blue). 
Red lines are fits to the curves, with exponents $\mu$ = -1.55 for {\em r} = 1000 and 
$\mu$ = -1.45 for {\em r} = 10000. Different exponents describe the power-law 
regime for different sampling ratios. The power-law 
persists long in time over different 
orders of magnitude.
Frame (b) shows the evolution of W(t)/t$^{-1.55}$ versus log(t) for various {\em r} 
and grids $W_{box}$ = 128000 (black) and $W_{box}$ = 10000 (other colours). 
The exponent $\mu$ = -1.55 is appropriate only for the case {\em r} = 1000 
where W(t)/t$^{-1.55}$ shows a constant mean value. 
}

\newpage
{
\baselineskip=1cm
Figure 5\\
We compare the different behaviours for $a = a_{\infty}$ (frame (a)) and 
$a = a_{tg}$ (frame (b)), when all the N initial conditions are randomly selected 
inside a single bin of the phase space partition. 
We used in both cases W$_{box}$ = 64000. 
Frame (a) illustrates the two 
different power-law coefficients obtained for $a = a_{\infty}$ when 
the initial conditions are uniformly spread all over the phase space or are 
constrained inside a single bin. 
In the frame we show four different selections of the initial bin in 
different colours. The four curves overlap in the regime of convergence to 
the attractor, oscillating at high frequency among two states every 
temporal step and giving rise 
to the broad coloured curves. 
In frame (b) the two different selections 
of initial conditions exhibit the same power-law exponent.
}

\end{document}